\documentclass[aps,prr,showpacs,footinbib,superscriptaddress,twocolumn,longbibliography]{revtex4-2}

\usepackage{amsfonts}
\usepackage{amsmath}
\usepackage{multirow}
\usepackage{txfonts}
\usepackage{amssymb}
\usepackage{amsbsy}
\usepackage{graphicx}
\usepackage{epstopdf}
\usepackage{color}
\usepackage{braket} 
\usepackage{mathdots} 
\usepackage{booktabs}
\usepackage{indentfirst}
\usepackage{hyperref}
\usepackage{diagbox}
\usepackage{lipsum}
\usepackage{mwe}
\hypersetup{ colorlinks=true, linkcolor=blue, anchorcolor=blue, citecolor=blue,urlcolor=blue}
\begin{document}
\title{Green's functions of multiband non-Hermitian systems}	
	
\author{Yu-Min Hu}
\affiliation{Institute for Advanced Study, Tsinghua University, Beijing,  100084, China}
\affiliation{Department of Physics, Princeton University, Princeton, New Jersey 08544, USA}
	
\author{Zhong Wang}
\affiliation{Institute for Advanced Study, Tsinghua University, Beijing,  100084, China}  
\date{\today}
\begin{abstract}
	Green's functions of non-Hermitian systems play a fundamental role in various dynamical processes. Because non-Hermitian systems are sensitive to boundary conditions due to the non-Hermitian skin effect, open-boundary Green's functions are closely related to the non-Bloch band theory. While the exact formula of open-boundary Green's functions in single-band non-Hermitian systems proves to be an integral along the generalized Brillouin zone (GBZ), the proper generalization in generic multiband systems remains unclear. In this paper, we derive a formula of open-boundary Green's functions in multiband non-Hermitian systems by viewing the multiband GBZ on the Riemann surface. This formula can be applied to describe directional amplification in multiband systems, which can be verified at various experimental platforms.
\end{abstract}
\maketitle
\section{Introduction} 
Numerous intriguing properties emerging in non-Hermitian systems have attracted increasing attention \cite{Ashida2020,Bergholtz2021RMP}. One of the fascinating non-Hermitian features is the non-Hermitian skin effect (NHSE), where bulk states under open boundary conditions (OBCs) accumulate at the boundaries  \cite{yao2018edge,yao2018chern,kunst2018biorthogonal,Alvarez2018non-hermitian,lee2018anatomy,xiao2020non,helbig2020generalized,Ghatak2019NHSE,Weidemann2020topological,wangwei2022non,Zhang2022ReivewOnNHSE,ding2022non,lin2023topological}. The NHSE invalidates the conventional Bloch-band picture and leads to the non-Bloch band theory based on the generalized Brillouin zone (GBZ) \cite{yao2018edge,Yokomizo2019}. Initially developed to elucidate non-Hermitian topology \cite{yao2018edge,yao2018chern,Song2019real,Yokomizo2019,kunst2018biorthogonal,Yang2020Auxiliary,Deng2019,Borgnia2019,Longhi2020chiral,Kawabata2020nonBloch,liu2019second,Lee2020Unraveling,Yi2020,wang2022amoeba}, the non-Bloch band theory also underlies various non-Hermitian dynamical phenomena \cite{Song2019,Longhi2019Probing,Longhi2019nonBloch,Xiao2021nonBloch,xue2022non,Longhi2022self-healing,longhi2022non}.
	
To study the responses to external perturbations in a non-Hermitian system with the NHSE, OBC Green's functions in one-dimensional (1D) single-band models and some simple multiband models have been formulated as a contour integral on the GBZ \cite{xue2021simple,Wanjura2019,Okuma2021non}. The GBZ-based formula can be easily evaluated by the residue theorem, the results of which have been applied to directional amplification \cite{xue2021simple} and quantized topological response \cite{li2021quantized}.
	
Despite the fundamental significance of non-Hermitian Green's functions, there is currently no proper generalization of GBZ-based formulas in generic multiband non-Hermitian systems. One of the difficulties is that energy bands are multivalued functions \cite{Li2020critical,Yang2020Auxiliary,fu2023anatomy,liu2023topological}. Additionally, a multiband GBZ has complex substructures denoted by sub-GBZs \cite{Yang2020Auxiliary}. As a result, multiband systems with complicated sub-GBZs exhibit many unique properties that are dramatically different from single-band models whose GBZ is a single loop. These differences complicate the direct generalization of GBZ-based Green's functions.

To address this problem, we develop the theory of non-Hermitian Green's functions in generic multiband systems from the perspective of the Riemann surface attached to the multivalued energy function. Let us consider a generic multiband non-Hermitian Bloch Hamiltonian $h(k)=\sum_{n=-m}^ma_ne^{ikn}$, where $a_{-m},\dots,a_m$ are $l\times l$ hopping matrices between different unit cells, with $l$ being the number of orbitals in a unit cell. The real-space Hamiltonian $H$ can be easily generated by $h(k)$. Our central result is the GBZ-based formula of the OBC Green's function $G(\omega)=\frac{1}{\omega-H}$ of multiband non-Hermitian systems:
\begin{eqnarray}
	\braket{x,a|G(\omega)|y,b}=\sum_{j=1}^l\int_{\text{GBZ}_j}\frac{\mathrm{d}\beta}{2\pi i\beta}\beta^{x-y}\frac{\braket{a|R_j,\beta}\braket{L_j,\beta|b}}{\omega-E_j(\beta)},\label{eq:result}
\end{eqnarray}
where $x,y$ are the spatial locations of different unit cells and $a,b$ represent the internal orbitals in each unit cell. In the above equation, with the definition $h(\beta)\equiv h(e^{ik}\to\beta)$, we have $h(\beta)\ket{R_j,\beta}=E_j(\beta)\ket{R_j,\beta}$ and $\bra{L_j,\beta}h(\beta)=\bra{L_j,\beta}E_j(\beta)$ for the general complex $\beta$. Here, $\ket{R_j,\beta}$ and $\ket{L_j,\beta}$ with $j=1,2,\dots,l$ are biorthogonal eigenstates of $h(\beta)$.  Also, $E_j(\beta)$ is the eigenvalue of $h(\beta)$ on the $j$th Riemann sheet of the Riemann surface. The Riemann surface consists of all the solutions $(\beta,E)$ of the algebraic equation $\det[E-h(\beta)]=0$,  which form a 1D complex subspace of the two-dimensional (2D) complex $(\beta,E)$ space.

We shall derive this formula as an integral along sub-GBZs [$\text{GBZ}_j$ in Eq. \eqref{eq:result}], which is viewed on the Riemann surface rather than simply on the complex $\beta$ plane. To do this, we first investigate the properties of the GBZ in multiband non-Hermitian systems. Concretely, on the Riemann surface determined by the characteristic equation $\det[E-h(\beta)]=0$, the multiband GBZ divides into several sub-GBZs that are attached to different Riemann sheets \cite{Yang2020Auxiliary}. These sub-GBZs are denoted by $\text{GBZ}_j$ with $j$ being the sheet index. Each sub-GBZ is associated with one non-Bloch band of the OBC spectrum. Crucially, the sub-GBZs form a boundary on the Riemann surface, separating the $\beta$ roots of the characteristic equation into two distinct groups. 

Based on these properties, we then formulate the multiband Green's functions by using contour deformation on each Riemann sheet. In the end, we obtain the GBZ-based formula Eq. \eqref{eq:result}, which provides asymptotically exponential behaviors that are analogous to those found in single-band systems or in simple multiband systems with overlapping sub-GBZs \cite{xue2021simple}. The formula is independent of artificial choices of frequency-dependent integral contours. It unveils the vital role of multiband sub-GBZs in non-Hermitian dynamics and elucidates the response properties of all energy bands in multiband systems. Given that Green's functions are closely related to experiments, our results are readily applicable to various experimental platforms where NHSE is observed \cite{Ghatak2019NHSE,Weidemann2020topological,helbig2020generalized,xiao2020non,wangwei2022non}.
	
This paper is organized as follows. In Sec. \ref{sec:GBZ}, we first discuss the basic facts of the multiband GBZ from the perspective of the Riemann surface. Next, in Sec. \ref{sec:geometric}, we elucidate that the multiband GBZ on the Riemann surface is a geometric boundary separating the roots of the characteristic equation. Then, in Sec. \ref{sec:GF}, we derive the formula of multiband non-Hermitian Green's functions as an integral on the sub-GBZs. Finally, we make several concluding remarks in Sec. \ref{sec:conclusion}.

\section{Multiband GBZ and Riemann surface}\label{sec:GBZ}

To begin with, we consider a general multiband non-Hermitian model and discuss its OBC band structure within the framework of non-Bloch band theory. Specifically, we elucidate that the multiband nature of the OBC spectrum can be effectively represented by a Riemann surface associated with the non-Hermitian system. From this point of view, the sub-GBZ of each OBC energy band lives on one of the Riemann sheets.

Without loss of generality, we consider a multiband model with $l$ energy bands. The $l\times l$ Bloch Hamiltonian is given by $h(k)=\sum_{n=-m}^ma_ne^{ikn} $, with $a_{-m},\dots,a_m$ being $2m+1$ hopping matrices of dimension $l\times l$. A Hermitian multiband model is obtained if $a_{-n}=a_n^\dagger$. In the non-Bloch band theory, it is convenient to define a non-Bloch Hamiltonian by replacing $e^{ik}$ with a complex number $\beta$:
\begin{equation}
	h(\beta)=\sum_{n=-m}^ma_n\beta^n.\label{eq:hbeta}
\end{equation}  
Immediately, the characteristic equation of the OBC eigenequation $H\ket{\psi}=E\ket{\psi}$ is given by
	\begin{eqnarray}
		\det[E-h(\beta)]=0.\label{eq:char}
	\end{eqnarray}  
This is an algebraic equation of both $\beta$ and $E$. With a specific $E$, there are $2ml$ roots of $\beta$ which are sorted by their norms: $|\beta_1(E)|\le|\beta_2(E)|\le\cdots\le|\beta_{2ml}(E)|$. For later convenience, we define $M=ml$.  The OBC spectrum of this multiband system can be determined by the non-Bloch band theory, where the GBZ is given by \cite{yao2018edge,Yokomizo2019}
	\begin{eqnarray}
		|\beta_M(E)|=|\beta_{M+1}(E)|.\label{eq:GBZ}
	\end{eqnarray}
This GBZ equation originates from the requirement that OBC wave functions should fulfill boundary conditions at two edges \cite{yao2018edge,Yokomizo2019}. With this equation, we are ready to obtain the OBC spectrum and the localization length of skin modes. 
		\begin{figure*}[t]
		\includegraphics[width=17cm]{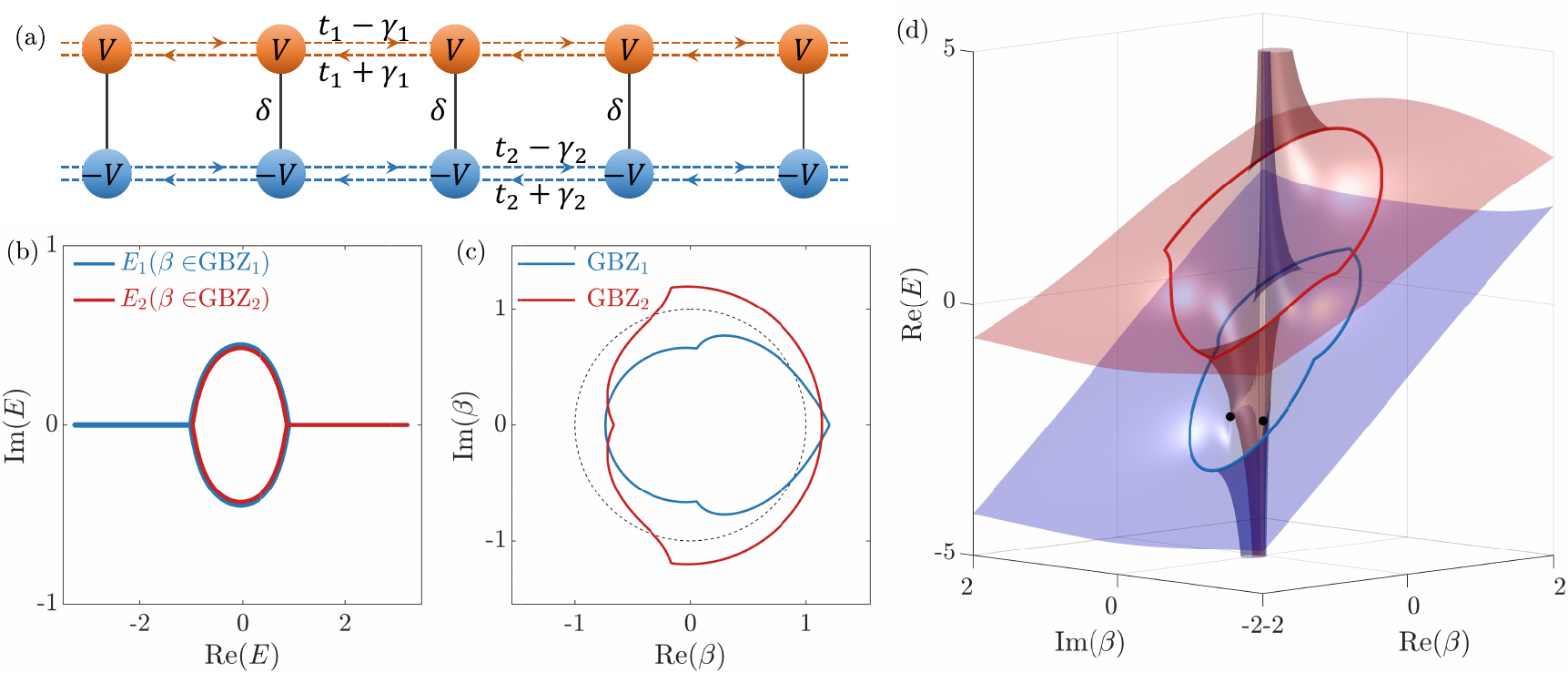}
		\caption{(a) The model. (b) Two non-Bloch energy bands under open boundary conditions. (c) The generalized Brillouin zone (GBZ) with two sub-GBZs. Red/blue lines in (b) and (c) represent the sheet indexes of the Riemann surface. (d) The Riemann surface of these two bands, with colored lines being sub-GBZs on each Riemann sheet. The two black points are the branch points (exceptional points) connecting two sheets. Parameters:  $t_1=1,\ \gamma_1=-0.3,\ t_2=1, \gamma_2=0.5,\ \delta=1,$ and $V=0.8$. We keep these parameters throughout this paper.}
		\label{fig:model}
	\end{figure*}
	
While the results in this paper are quite general in multiband non-Hermitian systems, we employ a particular model as an illustrative example:
	\begin{eqnarray}
		h(\beta)=\begin{pmatrix}(t_1+\gamma_1)\beta+\frac{t_1-\gamma_1}{\beta} +V&\delta \\ \delta &(t_2+\gamma_2)\beta+\frac{t_2-\gamma_2}{\beta}-V\end{pmatrix}.\label{eq:cHN}
	\end{eqnarray}
The real-space Hamiltonian is presented in Fig. \ref{fig:model}(a). This model describes two Hatano-Nelson models coupled by intracell hoppings. It was proposed to illustrate the critical NHSE when $\delta\to0$ \cite{Li2020critical,Liu2020helical,Yookomizo2021scaling}. If $\delta=0$, the two Hatano-Nelson chains will decouple, and we must separately apply the non-Bloch band theory to each chain. In this paper, we focus on the general choices of parameters with nonzero $\delta$ and develop its multiband OBC Green's functions below.  The characteristic equation $\det[E-h(\beta)]=0$ of this two-band model has four roots $|\beta_1(E)|\le|\beta_2(E)|\le|\beta_3(E)|\le|\beta_4(E)|$. Thus, the GBZ equation becomes $|\beta_2(E)|=|\beta_3(E)|$. A typical GBZ of this model is shown in Fig. \ref{fig:model}(c), which has substructures called sub-GBZs  \cite{Yang2020Auxiliary} (labeled by red/blue lines in the plot).
	
Interestingly, Eq. \eqref{eq:char} is also a high-order algebraic equation of the complex energy $E$. From this point of view, we can rewrite this equation as
	\begin{equation}
	\det[E-h(\beta)]=\prod_{j=1}^{l}[E-E_j(\beta)]=0\label{eq:sheet}
	\end{equation}
where $E_1(\beta),\ E_2(\beta),\ \dots\ ,\ E_l(\beta)$ are the solutions of $E$ at a given $\beta$. They are the $l$ Riemann sheets of the Riemann surface. As a direct application, we point out that, when $\beta$ goes around the unit circle, $E_j(e^{ik})$ provides the Bloch spectrum of the $j$th energy band under periodic boundary conditions. 
	
The collection of pairs $\{\beta,E_j(\beta)\}$, where $\beta$ is an arbitrary complex variable, constitutes the $j$th sheet of the Riemann surface. From this point of view, we observe that each energy band within a non-Hermitian multiband system finds its place on a distinct Riemann sheet \cite{Yang2020Auxiliary}. Furthermore, it is crucial to note that each band is accompanied by a corresponding sub-GBZ. For simplicity, we label the sub-GBZ on the $j$th sheet as $\text{GBZ}_j$. The non-Bloch band theory reveals that the complex variable $\beta$ moving along the $\text{GBZ}_j$ gives rise to the $j$th non-Bloch energy band $E_j(\beta)$ of the OBC spectrum (Fig. \ref{fig:model}). This is the non-Bloch generalization of the Bloch band structure. 
	
We consider the characteristic equation [Eq. \eqref{eq:sheet}]  for an arbitrary energy $E$. The characteristic equation has $2M$ roots $\beta_1(E),\dots,\beta_{2M}(E)$. At each root, $\tilde\beta\equiv\beta_k(E)$, with the root index $k$ running in $\{1,\dots,2M\}$, Eq. \eqref{eq:sheet} has $l$ different energy solutions $E_1(\tilde{\beta}),\dots,E_l(\tilde{\beta})$. Among these, there should be a unique $E_j(\tilde{\beta})$ that satisfies $E_j(\tilde{\beta})=E$, where we ignore accidental degeneracy. The correspondence between the energy $E$ and the sheet index $j$ indicates that the solution set $\{\beta_k(E),E\}$ stays on the $j$th Riemann sheet. In other words, this procedure attaches a unique sheet index $j$ to each $\beta$ root of Eq. \eqref{eq:char}. Therefore, we can label the root as $\beta_{k}^{(j)}(E)$, where $k$ is the root index, and $j$ is the sheet index determined by $k$. It is worth noting that, for the $2M$ roots of a specific $E$, the sheet index $j$ does not necessarily cover all elements in $\{1,2,\dots,l\}$ but the root index $k$ always runs from $1$ to $2M$. Furthermore, the GBZ equation [Eq. \eqref{eq:GBZ}] can be written as
	\begin{equation}
		|\beta_M^{(j)}(E)|=|\beta_{M+1}^{(j^\prime)}(E)|,\label{eq:sheetGBZ}
	\end{equation}
where $j$ and $j^\prime$ can take different sheet indexes. Indeed, although they share the same OBC energy, $\{\beta_M^{(j)}(E),E\}$ and $\{\beta_{M+1}^{(j^\prime)}(E),E\}$ satisfying the above equation can belong to different Riemann sheets, contributing to $\text{GBZ}_{j}$ and $\text{GBZ}_{j^\prime}$, respectively.

As an illustration, we apply the multiband GBZ theory to the model in Fig. \ref{fig:model}. The two-band model has two sub-GBZs, labeled by red and blue curves, respectively, in Fig. \ref{fig:model}(c). The two Riemann sheets are shown in Fig. \ref{fig:model}(d). We label the sheet with the blue (red) sub-GBZ as the first (second) sheet $E_1(\beta)$ [$E_2(\beta)$]. Similarly, the blue (red) sub-GBZ is denoted by $\text{GBZ}_{1}$ ($\text{GBZ}_2$). When $\beta$ goes around  $\text{GBZ}_{1}$ ($\text{GBZ}_2$), the OBC spectrum $E_1(\beta)$ [$E_2(\beta)$] in Fig. \ref{fig:model}(b) is shown by solid lines with the same colors. It is worth highlighting that there is an overlap between certain parts of the red and blue OBC spectra. When extracting energy values from this particular loop, the associated GBZ equation takes the form of Eq. \eqref{eq:sheetGBZ} where $j\ne j^\prime$. 

 \section{Geometric aspect of multiband GBZ}\label{sec:geometric}
After placing the multiband GBZ on the Riemann surface, we are now prepared to elucidate a significant geometric feature: the multiband GBZ is a boundary that distinguishes the roots of the characteristic equation when the energy $E$ lies outside the OBC spectrum. 
	
To see this point, we first review the results in single-band non-Hermitian systems \cite{Zhang2020correspondence,Okuma2020}. Given a typical single-band non-Bloch Hamiltonian $h(\beta)=\sum_{n=-M}^Mt_n\beta^n$ with $t_n$ being complex numbers, the characteristic equation $E-h(\beta)=0$ has $2M$ roots $|\beta_1(E)|\le\cdots\le|\beta_{2M}(E)|$ ordered by their norms. An elegant theorem states that the single-band GBZ is a boundary between the smallest $M$ roots $\{\beta_1(E),\beta_2(E),\dots,\beta_M(E)\}$ and the largest $M$ roots $\{\beta_{M+1}(E),\beta_{M+2}(E),\dots,\beta_{2M}(E)\}$ for arbitrary $E$. This theorem comes from the fact that the OBC spectrum has a vanishing winding number. That is, $W(E)\equiv\frac{1}{2\pi }\int_{\text{GBZ}}\frac{\mathrm{d}}{\mathrm{d}\beta}\log[E-h(\beta)]\mathrm{d}\beta=0$ when $E$ is not on the OBC spectrum. Since the polynomial $E-h(\beta)$ has an order-$M$ pole at the origin, there must be $M$ roots of $E-h(\beta)=0$ inside the GBZ. A more careful analysis shows that the GBZ encircles the smallest $M$ roots $\{\beta_1(E),\beta_2(E),\dots,\beta_M(E)\}$ while the other $M$ roots with larger norms stay outside. In the end, the single-band GBZ is a boundary between the smallest $M$ roots $\{\beta_1(E),\beta_2(E),\dots,\beta_M(E)\}$ and the largest $M$ roots $\{\beta_{M+1}(E),\beta_{M+2}(E),\dots,\beta_{2M}(E)\}$ when $E$ is not on the OBC spectrum.
	
Now we discuss the analogous property for the GBZ in multiband non-Hermitian systems. While the fact that different sub-GBZs belong to different sheets makes it subtle to consider winding numbers of sub-GBZs, we can bypass this subtlety via the following argument.

Starting from Eq. \eqref{eq:char}, we consider a large $E$ far away from the OBC spectrum ($|E|\to+\infty$). In this limit, the characteristic equation has $M$ roots $\{\beta_1^{(j_1)}(E),\dots,\beta_M^{(j_M)}(E)\}$ close to the origin. Meanwhile, the other $M$ roots $\{\beta_{M+1}^{(j_{M+1})}(E),\dots,\beta_{2M}^{(j_{2M})}(E)\}$ go to infinity, no matter what sheets they belong to. Consequently, all sub-GBZs on different Riemann sheets naturally separate these $2M$ points $\{\beta_k^{(j_k)}(E),E\} $ into two groups. 
	
Let us pull $E$ back from infinity. At the same time, these $2M$ roots $\{\beta_k^{(j_k)}(E),E\}$ will move on the Riemann surface. During this process, the sheet indexes $j$ of these roots may change when $E$ crosses some branch cuts on the Riemann surface. The order indexes $k$ may also exchange within $\{\beta_1^{(j_1)}(E),\dots,\beta_M^{(j_M)}(E)\}$ or $\{\beta_{M+1}^{(j_{M+1})}(E),\dots,\beta_{2M}^{(j_{2M})}(E)\}$. If the path of $E$ does not cross the OBC spectrum on the complex energy plane, any point $\{\beta_k^{(j_k)}(E),E\}$ on the $j_k$th sheet cannot cross the $\text{GBZ}_{j_k}$. Although a point $\{\beta_{k^\prime}^{(j_{k^\prime})}(E),E\}$ on the $j_{k^\prime}$th sheet may have the same $\beta$ as points on $\text{GBZ}_{j_k}$ with $j_k\ne j_{k^\prime}$, this point does not belong to $\text{GBZ}_{j_k}$. In other words, there is no path on the Riemann surface to move a $\beta$ from the neighbor of the origin to infinity without crossing sub-GBZs. Since this case is smoothly deformed from the aforementioned case $|E|\to\infty$, we conclude that all sub-GBZs together separate the roots of the characteristic equation into two distinct groups: the smallest $M$ roots $\{\beta_1^{(j_1)}(E),\dots,\beta_M^{(j_M)}(E)\}$ and the largest $M$ roots $\{\beta_{M+1}^{(j_{M+1})}(E),\dots,\beta_{2M}^{(j_{2M})}(E)\}$.

When the deformation path of $E$ crosses the OBC spectrum, there must be a point $\{\beta_M^{(j_M)}(E),E\}$ on the $j_M$th sheet moving out of $\text{GBZ}_{j_M}$ while another point $\{\beta_{M+1}^{(j_{M+1})}(E),E\}$ on the $j_{M+1}$th sheet moves into $\text{GBZ}_{j_{M+1}}$. Then they exchange their root indexes $M$ and $M+1$. Eventually, all sub-GBZs on the Riemann surface still contain $\{\beta_1^{(j_1)}(E),\dots,\beta_M^{(j_M)}(E)\}$ inside. In conclusion, the sub-GBZs on the Riemann surface form a boundary separating two groups of $\beta$ roots of the characteristic equation Eq. \eqref{eq:char}.
	
We emphasize that the sub-GBZs are not a natural boundary if we view them as close curves on the complex $\beta$ plane. Whereas the sub-GBZs may intersect with each other when viewed on the complex $\beta$ plane, they belong to different Riemann sheets and therefore are disconnected from the perspective of the Riemann surface. 

\begin{figure*}[t]
		\centering
		\includegraphics[width=\textwidth]{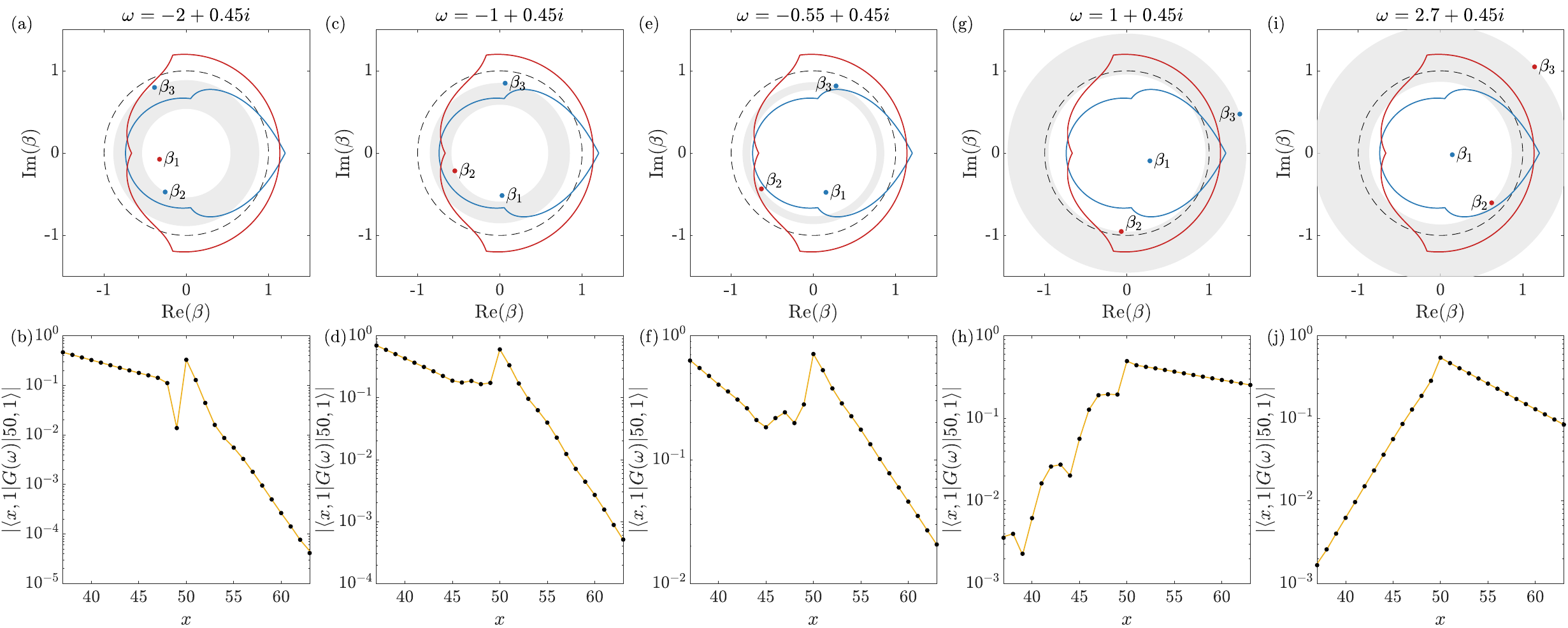}
		\caption{The upper row shows generalized Brillouin zones (GBZs) and the distribution of roots of $\det[\omega-h(\beta)]=0$. Different colors of roots and sub-GBZs represent their sheet indexes. The fourth root is outside the plot. The circular gray region is $|\beta_{M}(\omega)|<R<|\beta_{M+1}(\omega)|$. The lower row displays the multiband Green's functions $\braket{x,1|G(\omega)|50,1}$, where the index $1$ means the first orbital in the unit cell. Black dots come from numerical results on an open chain with $L=100$; yellow lines are obtained by integrating Eq. \eqref{eq:result}. Each column is labeled by a different frequency shown above. }\label{fig:GF}
	\end{figure*}

For example, in Fig. \ref{fig:GF}(a) and (c), it seems that the red sub-GBZ encloses $\beta_3$ when viewed on the complex $\beta$ plane. However, the blue point $\beta_3$ belongs to the other sheet and stays outside of the blue sub-GBZ. This is fundamentally different from the situation in single-band models where there is only a single energy sheet and, consequently, a single GBZ loop. In multiband non-Hermitian systems, it is more natural to discuss the root distributions on the Riemann surface, instead of the complex $\beta$ plane.
	
From the perspective of the Riemann surface, we conclude that all sub-GBZs form a natural boundary dividing $\{\beta_1^{(j_1)}(E),\dots,\beta_M^{(j_M)}(E)\}$ and $\{\beta_{M+1}^{(j_{M+1})}(E),\dots,\beta_{2M}^{(j_{2M})}(E)\}$ for arbitrary $E$ not being on the OBC spectrum. We will use this fact to formulate multiband non-Hermitian Green's functions in the next section.
\section{Multiband non-Hermitian Green's functions}\label{sec:GF}

Based on the last section, we can now construct Green's functions for multiband non-Hermitian systems. Considering a real-space OBC Hamiltonian $H$ generated by Eq. \eqref{eq:hbeta}, the Green's function is defined as
	\begin{eqnarray}
		\braket{x,a|G(\omega)|y,b}=\braket{x,a|\frac{1}{\omega-H}|y,b},
	\end{eqnarray}
where $x,y$ are the spatial locations of different unit cells and $a,b$ are the internal orbitals in each unit cell. For convenience, we take a complex frequency $\omega$, whose imaginary part is interpreted as the total loss or gain added to the OBC Hamiltonian. Additionally, $\omega$ is not taken from the OBC spectrum.

	\subsection{Green's functions as a contour integral}
 
We shall make a connection between the OBC Green's function and an integral on the multiband GBZ. To do this, we first show that the OBC Green's function can be expressed as the following contour integral:
	\begin{eqnarray}
		\braket{x,a|G(\omega)|y,b}=\int_{|\beta|=R}\frac{\mathrm{d}\beta}{2\pi i\beta}\beta^{x-y}\braket{a|\frac{1}{\omega-h(\beta)}|b}.\label{eq:Rint}
	\end{eqnarray}
The integral contour is a circle with a radius $R$, which should satisfy $|\beta_M(\omega)|<R<|\beta_{M+1}(\omega)|$, with $\beta_M(\omega)$ and $\beta_{M+1}(\omega)$ being the middle two roots of $\det[\omega-h(\beta)]=0$ (gray region in Fig. \ref{fig:GF}). It is important to note that this integral contour is not equivalent to the conventional Brillouin zone $|\beta|=1$. 

Why is this circular region so special? To answer this question, we use the language of Teoplitz matrices and generalize the argument in Ref. \cite{xue2021simple} into multiband systems. A more formal proof can be found in Ref. \cite{wang2022amoeba}.

Given a matrix-valued Laurent polynomial $f(\beta)=\sum_nc_n\beta^n$ with $c_n$ being $l\times l$ coefficient matrices, a block Teoplitz matrix $T(f)$ is defined as $T_{jk}(f)=c_{k-j}=\int_{|\beta|=R}\frac{\mathrm{d}\beta}{2\pi i\beta}\beta^{j-k}f(\beta)$ for an arbitrary $R$. We denote an $l\times l$ block of the whole Teoplitz matrix as $ T_{jk}(f)$. In other words, $T(f)$ is expressed as the following block structure:
\begin{eqnarray}
	T(f)=\begin{pmatrix}
		c_0 &c_1 &c_2& \dots&\\
		c_{-1}&c_0 &c_1&\ddots\\
		c_{-2}&	c_{-1}&c_0 &\ddots\\
		\vdots &\ddots&\ddots&\ddots
	\end{pmatrix}.
\end{eqnarray}
This is just a real-space Hamiltonian if we interpret $f(\beta)$ as a non-Bloch Hamiltonian $h(\beta)$ in Eq. \eqref{eq:hbeta}. The OBC Hamiltonian $H$ is obtained by truncating $T(h)$ as an $Ll\times Ll$ matrix with $L$ being the number of unit cells.

Now let us take two block Teoplitz matrices $T(f_1)$ and $T(f_2)$ generated by two $l\times l$ matrix-valued Laurent polynomials $f_1(\beta)=\sum_na_n\beta^n$ and  $f_2(\beta)=\sum_nb_n\beta^n$ respectively. There is a simple identity $T(f_1)T(f_2)=T(f_1f_2)$ coming from $\sum_k T_{i k}(f_1) T_{k j}(f_2)=\sum_k a_{k-i} b_{j-k}=(f_1f_2)_{j-i}=T_{i j}(f_1f_2)$. Strictly speaking, the identity is an approximation without considering the boundary contributions. Such a boundary correction is exponentially small in the bulk compared with $(f_1f_2)_{j-i}$. Hence, we take the thermodynamic limit and ignore the boundary effect. An immediate consequence of the identity is that $T(f)T(f^{-1})=1$, namely, $[T(f)]^{-1}=T(f^{-1})$. 

To get the OBC Green's functions, we replace $f(\beta)$ with $\omega-h(\beta)$ to obtain the expression $\{T[\omega-h(\beta)]\}^{-1}=T[\frac{1}{\omega-h(\beta)}]$. The matrix elements $\braket{x,a|\{T[\omega-h(\beta)]\}^{-1}|y,b}$ are given by the Laurent expansion of $\braket{a|\frac{1}{\omega-h(\beta)}|b}$, as shown in Eq. \eqref{eq:Rint}. Because $\frac{1}{\omega-h(\beta)}$ has poles at the roots of $\det[\omega-h(\beta)]=0$, our next step is to specify which $R$ is proper for the Laurent expansion. 

Consider a smooth interpolation $f_{s}(\beta)$ where $f_{s=1}(\beta)=f(\beta)$ and $f_{s=0}(\beta)=\mathbb{I}_{l\times l}$. The Teoplitz matrix generated by $f_{s=0}(\beta)=\mathbb{I}_{l\times l}$ trivially satisfies $T(f_{s=0})T(f_{s=0}^{-1})=1$. Then the smooth interpolation of $[f_{s}(\beta)]^{-1}$ from $s=0$ to $1$ provides the proper expansion of $[f_{s=1}(\beta)]^{-1}$. This requires that $f_{s}(\beta)\ne0$ on the integral circle $|\beta|=R$. The topological winding number $W_s=\frac{1}{2\pi}\int_{|\beta|=R}\mathrm{d}\ln\{\det[f_s(\beta)]\}=0$ remains unchanged during smooth interpolation because it has a vanishing value at $s=0$.  Now we take $f_{s=1}(\beta)=\omega-h(\beta)$ to discuss multiband non-Hermitian Green's functions. Considering that $\det[\omega-h(\beta)]=\frac{C_M}{\beta^M}\prod_{k=1}^{2M}[\beta-\beta_k(\omega)]$ with $\beta_{k=1,\dots,2M}(\omega)$ being $2M$ roots ordered by their norms, the requirement of a zero winding number leads to radius $R$ satisfying $|\beta_M(\omega)|<R<|\beta_{M+1}(\omega)|$. In the end, we get back to the integral form of Eq. \eqref{eq:Rint}.

\subsection{GBZ-based formula}

The above discussion indicates that the allowed values of $R$ in Eq. \eqref{eq:Rint} depend on $\omega$, which is a shortcoming of the formula. To use this formula, we need to first specify the $\omega$-dependent $R$ before calculating the integral. It is more natural to find a unique integral contour that is independent of $\omega$. The work has been done in single-band non-Hermitian systems and some simple multiband systems with overlapping sub-GBZs. In these systems, it has been proved that OBC Green's functions are obtained from a contour integral on the GBZ \cite{xue2021simple}. Consequently, we expect that a similar integral exists in non-Hermitian multiband systems, where the multiband GBZ plays an indispensable role.

One possible way for the generalization is that we may substitute the integral contour $|\beta|=R$ in Eq. \eqref{eq:Rint} by a specific sub-GBZ. However, this is problematic (see Appendix \ref{appe:failure} for a detailed discussion). This is because non-Hermitian Green's functions describe the response of the whole system to external perturbations, instead of just one of the energy bands. Therefore, we expect to obtain a formula that considers all sub-GBZs.
		
A natural way to relate the circle $|\beta|=R$ to the sub-GBZs is to continuously deform the integral contour on the Riemann surface. To see this, we express Eq. \eqref{eq:Rint} as
	\begin{eqnarray}
		\braket{x,a|G(\omega)|y,b}=\int_{|\beta|=R}\frac{\mathrm{d}\beta}{2\pi i\beta}\beta^{x-y}\sum_{j=1}^l\frac{\braket{a|R_j,\beta}\braket{L_j,\beta|b}}{\omega-E_j(\beta)}.\label{eq:Proj_int}
	\end{eqnarray}
In the above equation, $h(\beta)\ket{R_j,\beta}=E_j(\beta)\ket{R_j,\beta}$ and $\bra{L_j,\beta}h(\beta)=\bra{L_j,\beta}E_j(\beta)$. Here, $E_j(\beta)$ is the eigenvalue of $h(\beta)$ on the $j$th Riemann sheet. The left and right eigenvectors constitute the biorthogonal basis with $\braket{L_j,\beta|R_{j^\prime},\beta}=\delta_{j,j^\prime}$ and $\sum_{j=1}^l\ket{R_j,\beta}\bra{L_j,\beta}=\mathbb{I}_{l\times l}$. For later convenience, we define $P_{j,ab}(\beta)=\braket{a|R_j,\beta}\braket{L_j,\beta|b}$, which is the matrix element of the projection operators. Note that the integrand $\frac{P_{j,ab}(\beta)}{\omega-E_j(\beta)}$ is defined on the $j$th sheet of the Riemann surface. Thus, the poles on other sheets will not contribute to the integral on the $j$th sheet.
	
In Sec. \ref{sec:geometric}, we already learn that sub-GBZs form a boundary that separates $\{\beta_1(\omega),\dots,\beta_{M}(\omega)\}$ and $\{\beta_{M+1}(\omega),\dots,\beta_{2M}(\omega)\}$ on the Riemann surface. The same is true for the circle $|\beta|=R$ with $|\beta_M(\omega)|<R<|\beta_{M+1}(\omega)|$, if we view this circle as a collection of $l$ identical copies on different sheets of the Riemann surface. Now we can freely deform the integral contour continuously from the $|\beta|=R$ loop to the corresponding sub-GBZ on the same sheet. Namely, $	\int_{|\beta|=R}\frac{\mathrm{d}\beta}{2\pi i\beta}\beta^{x-y}\frac{P_{j,ab}(\beta)}{\omega-E_j(\beta)} \to \int_{\text{GBZ}_j}\frac{\mathrm{d}\beta}{2\pi i\beta}\beta^{x-y}\frac{P_{j,ab}(\beta)}{\omega-E_j(\beta)}$ . The integral contour will not pass any poles when we track the deformation process on each Riemann sheet.
	
For example, as shown in Fig. \ref{fig:GF}, if we start from a circle in the gray region, where $|\beta_M(\omega)|<R<|\beta_{M+1}(\omega)|$, deforming this loop to the red sub-GBZ will not pass any poles on the same sheet labeled by red points. While this deformation may cross the blue poles on the other sheet [Figs. \ref{fig:GF}(a), \ref{fig:GF}(c), and \ref{fig:GF}(e)], these blue poles will not contribute to the integral on the sheet with the red sub-GBZ. The same argument is true for the deformation into the blue sub-GBZ, with examples in Figs. \ref{fig:GF}(e) and \ref{fig:GF}(g). 
	
In the end, the OBC Green's functions of multiband non-Hermitian systems can be expressed as an integral on all sub-GBZs, as shown in Eq. \eqref{eq:result}, which is a central result of this paper. Remarkably, unlike Eq. \eqref{eq:Rint},  this formula is free of artificially choosing $\omega$-dependent integral contours. This formula depends only on all sub-GBZs, unveiling that sub-GBZs play a fundamental role in the dynamics of multiband non-Hermitian systems. 

The formula Eq. \eqref{eq:result} goes back to the integral on the conventional Brillouin zone if the underlying system is Hermitian. It also naturally goes back to the single-band models \cite{xue2021simple}. If all the sub-GBZs of some simple multiband systems are overlapping on the $\beta$ plane, the formula in Eq. \eqref{eq:result} goes back to Eq. \eqref{eq:Rint} via replacing $R$ by the coincident GBZ \cite{xue2021simple}.

\subsection{Asymptotic behaviors}

To facilitate applications of the formula Eq. \eqref{eq:result}, it is easy to show that the OBC Green's functions exhibit the asymptotic behaviors:
	\begin{equation}
		\braket{x,a|G(\omega)|y,b}\sim\left\{\begin{aligned}
			&P_{j_{M},ab}[\beta_M(\omega)][\beta_M(\omega)]^{x-y}, &x\gg y;\\
			&P_{j_{M+1},ab}[\beta_{M+1}(\omega)][\beta_{M+1}(\omega)]^{-|x-y|},  &x\ll y.		\end{aligned}\right.
   \label{eq:asymptotic}
	\end{equation}

\begin{figure}
	\centering
	\includegraphics[width=8.5cm]{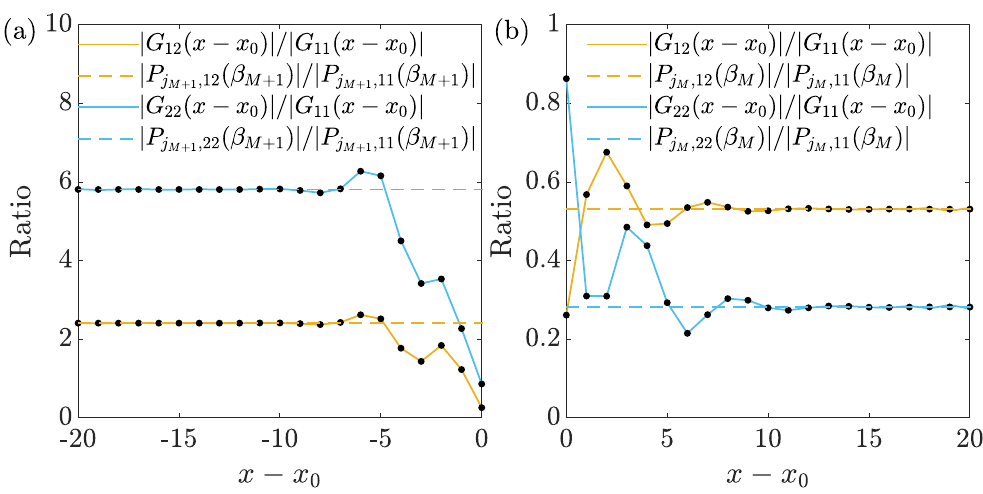}
	\caption{The ratios between matrix elements of multiband Green's functions. We fix $\omega=-0.55+0.45i$. Black points are numerical results of an $L=200$ chain, and we take $x_0=L/2$. We define $G_{ab}(x-x_0)=\braket{x,a|\frac{1}{\omega-H}|x_0,b}$ for the fixed frequency. The matrix elements of the $j_k$th projection operator at $\beta^{(j_k)}_k(\omega)$ are $P_{j_k,ab}(\beta_k)=\braket{a|R_{j_k},\beta_k}\braket{L_{j_k},\beta_k|b}$ with (a) $k=M+1$ and (b) $k=M$. }\label{fig:ratio}
\end{figure}

This is the multiband generalization of the asymptotic behaviors of single-band Green's functions \cite{xue2021simple}. Notably, in addition to the asymptotically exponential behaviors, our formula also provides the ratios between different matrix elements of the multiband Green's function. Concretely, the ratios between $\braket{x,a|G(\omega)|y,b}$ with fixed $x,y$ are given by the matrix elements of the projection operator $P_{j_k,ab}[\beta_k(\omega)]=\braket{a|R_{j_k},\beta_k(\omega)}\braket{L_{j_k},\beta_k(\omega)|b}$ at the pole $\beta_k(\omega)$ with $k=M+1$ if $x\ll y$ and $k=M$  if $x\gg y$. As defined in Sec. \ref{sec:GBZ}, $j_k$ is the sheet index of $\{\beta_k(\omega),\omega\}$. The numerical results in Fig. \ref{fig:ratio} show excellent agreement between the ratios $\frac{\braket{x,a|G(\omega)|y,b}}{\braket{x,a'|G(\omega)|y,b'}}$ and $\frac{P_{j_{M+1},ab}[\beta_{M+1}(\omega)]}{P_{j_{M+1},a'b'}[\beta_{M+1}(\omega)]}$ when $x\ll y$ (or $\frac{P_{j_{M},ab}[\beta_{M}(\omega)]}{P_{j_{M},a'b'}[\beta_{M}(\omega)]}$ when $x\gg y$). Therefore, our formula faithfully resolves the fine structures of multiband Green's functions between different orbitals. These fine structures cannot be seen from the formula Eq. \eqref{eq:Rint}.

\section{Discussions}\label{sec:conclusion}
We study the multiband GBZ from the perspective of the Riemann surface and propose a proper formula for OBC Green's functions in multiband non-Hermitian systems. The exact formula Eq. \eqref{eq:result} is an integral on the Riemann surface, with the integral contours being the sub-GBZs on the Riemann sheets. These contours are independent of the frequency in the Green's functions. In fact, they are intrinsic geometrical constructions of multiband non-Hermitian systems. The formula explicitly provides not only the asymptotic behaviors at a long distance but also the ratios between different orbitals.

The GBZ-based formula of multiband Green's functions has significant implications in various fields of non-Hermitian physics. From an experimental point of view, a variety of phenomena related to the NHSE have already been observed in experiments \cite{Ghatak2019NHSE,Weidemann2020topological,helbig2020generalized,xiao2020non,wangwei2022non}. Our formula can provide insights into the dynamical responses within these experimental systems. Furthermore, like its single-band counterparts, the asymptotic behaviors in Eq. \eqref{eq:asymptotic} offer a simple approach to predict the directional amplification of input signals in multiband bosonic open quantum systems  \cite{Wanjura2019,xue2021simple}. Meanwhile, the exponential growth of non-Hermitian Green's functions can also serve to detect nontrivial spectral winding in multiband non-Hermitian systems \cite{li2021quantized, Zirnstein2021bulk, Zirnstein2021exponential}.

From a theoretical perspective, it is known for Hermitian systems that the topological properties of multiband systems can be effectively characterized by Green's functions \cite{wang2012a}. Therefore, one can expect that the GBZ-based formula has potential applications in the non-Bloch bulk-boundary correspondence of multiband non-Hermitian systems. Another interesting direction is to extend the non-Hermitian Green's functions to higher-dimensional systems, for which the recently developed amoeba formulation could be useful \cite{wang2022amoeba}. Furthermore, it is also intriguing to investigate Green's functions for many-body non-Hermitian systems \cite{sen2020emergent, leeeunwoo2020many-body, lee2021many-body, kawabata2022many, alsallom2022fate}.

\section*{Acknowledgment}
We thank Fei Song and He-Ran Wang for helpful discussions and for their comments on the paper. This paper is supported by NSFC under Grant No. 12125405.
\appendix

\section{The Failure of the formulas based on a specific sub-GBZ}\label{appe:failure}

In the main text, we mention that substituting the integral contour $|\beta|=R$ in Eq. \eqref{eq:Rint} by a specific sub-GBZ is problematic. According to this approach, the Green's function formula would be an integral on $\text{GBZ}_j$ for a certain $j$ \footnote{Recently, the formula based on a specific sub-GBZ is adopted in Ref. \cite{fu2023anatomy}.}:
\begin{eqnarray}
	\braket{x,a|G(\omega)|y,b}\overset{\text{?}}{=}\int_{\text{GBZ}_j}\frac{\mathrm{d}\beta}{2\pi i\beta}\beta^{x-y}\braket{a|\frac{1}{\omega-h(\beta)}|b}.\label{apeq:subGBZint}
\end{eqnarray}	

\begin{figure}
	\centering
	\includegraphics[width=8.5cm]{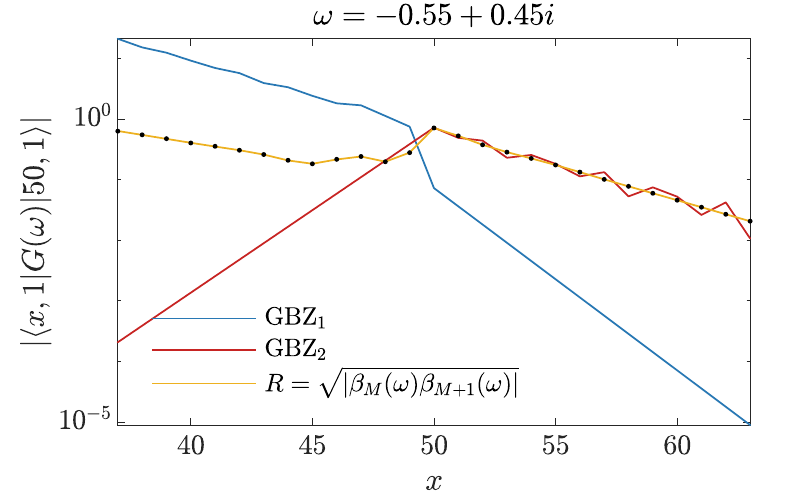}
	\caption{The failure of the formulas based on a specific sub-generalized Brillouin zone (sub-GBZ). The index $1$ in $\braket{x,1|G(\omega)|50,1}$ labels the first orbital in the unit cell. The yellow line is calculated by Eq. \eqref{eq:Rint} with a circle contour $R=\sqrt{|\beta_M(\omega)\beta_{M+1}(\omega)|}$, which is in excellent agreement with numerical results from an $L=100$ chain. The blue (red) line comes from Eq. \eqref{apeq:subGBZint} by choosing $\text{GBZ}_1$ ($\text{GBZ}_2$) to be the integral contour. The artificial choice of sub-GBZs fails to predict non-Hermitian Green's functions.}\label{fig:mismatch}
\end{figure}

In this appendix, we demonstrate the failure of this plausible formula in more detail. The results in Fig. \ref{fig:mismatch} show that integrating along a specific sub-GBZ does not predict the numerical values of multiband non-Hermitian Green's functions. This mismatch can be explained by contour deformation on the complex $\beta$ plane. 

As shown in Fig. \ref{fig:GF}, $|\beta|=R$ lies in the gray circular region between $|\beta_2(\omega)|$ and $|\beta_3(\omega)|$. Deforming $|\beta|=R$ on the complex $\beta$ plane into the red sub-GBZ in Figs. \ref{fig:GF}(a) and  \ref{fig:GF}(c) or into the blue sub-GBZ in Fig. \ref{fig:GF}(g) will inevitably cross the poles of $\frac{1}{\omega-h(\beta)}$ at $\beta_3(\omega)$ or $\beta_2(\omega)$, respectively. This process changes the integral results of Eq. \eqref{apeq:subGBZint}. The situation is even worse in Fig. \ref{fig:GF}(e) (the same parameter as Fig. \ref{fig:mismatch}), where the deformation to the red or blue sub-GBZ will always cross poles. Then Eq. \eqref{apeq:subGBZint}, with artificially choosing one of the sub-GBZs, will yield problematic results. 

A more physical reason for the failure of the above extension is that non-Hermitian Green's functions describe the response of the whole system to external perturbations, instead of just one of the energy bands. Therefore, the formula based on a specific sub-GBZ fails in predicting the Green's functions in general multiband non-Hermitian systems. 

The formula Eq. \eqref{apeq:subGBZint} is only applicable to special cases such as Fig. \ref{fig:GF}(i). In these cases, there is no pole between different sub-GBZs viewed on the complex $\beta$ plane. Hence, all sub-GBZs are equivalent to each other when we calculate the integral along them.

\bibliography{dirac}

\end{document}